\documentclass[12pt]{iopart}
\usepackage{graphicx}
\expandafter\let\csname equation*\endcsname\relax
\expandafter\let\csname endequation*\endcsname\relax
\usepackage{amsmath}
\usepackage{caption}
\usepackage{amsthm,amssymb,enumerate}
\usepackage{mathtools}
\usepackage{color}
\usepackage{bm}
\usepackage{paracol}
\newcommand{\grvec}[1]{\bm{#1}}
\newcommand{\matr}[1]{\bm{#1}}
\newcommand{\mS}{\matr{S}}
\newcommand{\mTheta}{\matr{\Theta}}
\renewcommand{\vec}[1]{\bm{#1}}

\newcommand{\llrrbrace}[1]{
	 \left\{\mkern-6mu\left\{#1\right\}\mkern-6mu\right\}}
\newcommand{\llrrbracket}[1]{
	 \left[\mkern-2mu\left[#1\right]\mkern-2mu\right]}

\newcommand{\llbracket}{ \left[ \mkern-2mu\left[ }
\newcommand{\rrbracket}{ \right] \mkern-2mu \right]}

\def\softd{{\leavevmode\setbox1=\hbox{d}%
         \hbox to 1.05\wd1{d\kern-0.4ex{\char039}\hss}}}

\begin{document}

\title[Shear-thinning in Oligomer Melts - Molecular Origins and Applications]{Shear-thinning in Oligomer Melts - Molecular Origins and Hybrid Multiscale Simulations}

\author{Ranajay Datta$^1$, Leonid Yelash$^2$,
Friederike Schmid$^1$, Florian Kummer$^3$, Martin Oberlack$^3$, M\'aria Luk\'a\v{c}ov\'a-Medvi{\softd}ov\'a$^2$, Peter Virnau$^1$}

\address{$^1$ Institute of Physics, Johannes Gutenberg University, Staudingerweg 9, 55128 Mainz, Germany}
\address{$^2$ Institute of Mathematics, Johannes Gutenberg University, Staudingerweg 9, 55128 Mainz, Germany}

\address{$^3$ Department of Mechanical Engineering, Technische Universität Darmstadt, Otto-Berndt-Str. 2, 64287 Darmstadt, Germany}

\ead{virnau@uni-mainz.de,lukacova@mathematik.uni-mainz.de}

\begin{abstract}
We investigate the molecular origin of shear-thinning in melts of flexible, semiflexible and rigid oligomers with coarse-grained simulations of a sheared melt.
Alignment, stretching and tumbling modes or suppression of the latter all contribute to understanding how macroscopic flow properties emerge from the molecular level. By performing simulations of single chains in a shear flow, we identify which of these phenomena are of collective nature and arise through interchain interactions and which are already present in dilute systems.
Building upon these microscopic simulations we identify 
by means of the Irving-Kirkwood formula the corresponding macroscopic stress tensor for a non-Newtonian polymer fluid. Shear-thinning effects in oligomer melts are also demonstrated by macroscopic simulations of a channel flow. The latter have been obtained by the discontinuous Galerkin method approximating macroscopic polymer flows.  Our study confirms the influence of microscopic details in the molecular structure of short polymers such as chain flexibility on macroscopic polymer flows.
\end{abstract}


\maketitle
\normalsize

\section{Introduction}
\label{sec_1}
Understanding the relation between viscosity and structure and its implications on macroscopic flow is of prime importance, particularly for semi-flexible polymers which are omnipresent in nature (DNA, actin filaments, microtubules) and synthetic polymers (polyelectrolytes, dendronized polymers). Technological applications are manifold and include e.g., purifying DNA in microfluidic devices \cite{Han_2000,Muthukumar_book,Kim_2017} or separation of polymers \cite{Weiss_2019}. Computer simulations at multiple scales nowadays provide powerful tools to probe and understand fundamental structure-flow relations as well as their applications. 

At the microscopic level non-equilibrium molecular dynamics (NEMD) simulations have been applied for forty years \cite{Kroger_1993,Chynoweth_1994,Kroger_1997,Moore_2000,Bair_2002,Yamamoto_2002,Celani2005,Huang2012,Xu_2017,Kong2019,Xu2020} to study how macroscopic flow properties such as shear thinning emerge from dynamics of microscopic structures. From the beginning fundamental issues were approached from two sides. On the one hand, generic bead-spring models (similar to the ones applied in this work) were used to probe the general behaviour of polymer melts under shear such as shear viscosity at zero shear rate \cite{Kroger_1993}, progressive alignment and elongation with shear and corresponding correlations with stress \cite{Kroger_1997}, the increase of viscosity when approaching the glass-transition in polymer melts and movement modes of individual chains \cite{Yamamoto_2002}. On the other hand, research in this direction was also driven by modeling and comparing simulations with actual experiments involving specific polymers \cite{Chynoweth_1994,Moore_2000,Bair_2002}. The underlying models tend to be more involved in these cases and include bending and even torsion terms absent in the early studies mentioned above. It therefore comes a bit as a surprise that the specific influence of stiffness on shear thinning in polymers has come only into the focus of attention, recently. Particularly, Ref.~\cite{Xu_2017} provides a comprehensive study and already anticipates some of the effects also discussed in this work, while \cite{Kong2019} focuses on the influence of chain stiffness on individual movement modes.

For complex multiscale systems bridging over large range of dynamically coupled scales is a challenging problem. In the last decades this question led to 
the development of new mathematical algorithms and hybrid multiscale methods.
One possibility to build a multiscale algorithm relies on the Lagrangian-Eulerian decomposition where the Lagrangian-type particles are embedded in the Eulerian description of fluid, see, e.g., \cite{emamy_2017, Ren_2005, Yasuda_2010}. Another approach  is based on the domain decomposition. Hereby a small accurate atomistic region is embedded into a coarser macrosopic model, see, e.g., \cite{Fedosov_2009}. Several hybrid models combining particle dynamics with the macroscopic continuum model can be found in the literature. In this context we should mention, e.g., the hybrid heterogeneous multiscale methods \cite{Weinan2007, E2003, E2004, E_2005, Ren_2007, Ren_2005, Yasuda_2010}, the seamless multiscale methods \cite{E2007,e_2007},
 the equation-free multiscale methods \cite{Kevrekidis_2003,Kevrekidis_2009}, the triple-decker atomistic-mesoscopic-continuum method \cite{Fedosov_2009}, or the internal-flow multiscale method \cite{lockerby2, lockerby1}. 
 A nice overview of multiscale flow simulations using particles is presented in \cite{koumoutsakos}.

In this paper we apply a hybrid multiscale method that couples atomistic details obtained by Molecular Dynamics with a continuum model approximated by the discontinuous Galerkin method.  In order to extract mean flow field information
from the Molecular Dynamics  averaging needs to
be performed. Specifically, the required rheological 
information for the complex stress tensor
is calculated by means of the Irving-Kirkwood formula \cite{Irving_1950}. Consequently, the averaged stress tensor is 
passed to the macroscopic continuum model.
Thus, our method belongs to the class of
hybrid particle-continuum methods under the statistical influence of microscale effects.  We note by passing that the degree of scale separation of a physical system influences the  sensitivity of the accuracy of a solution and the computational speed-up over a full molecular simulation~\cite{lockerby1}.

The present paper is organized in the following way. First, we review and explore with NEMD simulations the molecular foundation of shear-thinning in low molecular weight polymers as a function of chain stiffness in the framework of a microscopic bead-spring model (Section~\ref{sec_2} and \ref{sec_3}). In particular, we will show how shear-thinning emerges from an intricate interplay of molecular alignment, stretching and tumbling modes. The comparison of our high density melt with simulations of single chains will also allow us to provide educated guesses for the density dependence of individual effects. In Sections~\ref{sec_4} and \ref{sec_5} we use microscopic data obtained from simulations in Section~\ref{sec_3} as input for the study of various macroscopic channel flows using a hybrid multiscale method and investigate differences from Newtonian flow behaviour arising due to shear-thinning effects.

\section{Microscopic Model and Simulation Techniques}
\label{sec_2}
For our microscopic model we use standard bead-spring chains to represent the oligomers, as formulated by Kremer and Grest \cite{Kremer1990}. In this model, all beads interact with each other via a repulsive Weeks-Chandler-Andersen potential \cite{Weeks_1971}:

\begin{equation}
\begin{split}
V_{WCA}(r)&=4\epsilon\left [\left( \frac{\sigma}{r}\right) ^{12}-\left(\frac{\sigma}{r}\right)^{6} + \frac{1}{4}\right],\hspace{1cm}r<2^\frac{1}{6}\sigma \\
&=0,\hspace{4.62cm}r>2^\frac{1}{6}\sigma
\end{split}
\end{equation}
with $\sigma=1$ and $\epsilon=1$. Adjacent beads are connected with an additional FENE interaction \cite{FENE}: 

\begin{equation}
    V_{FENE}=-\frac{1}{2}KR^{2}\ln\left[1-\left(\frac{r}{R}\right)^{2}\right]
\end{equation}
with $K=30$ and $R=1.5$. Semiflexibility is implemented with a bending potential:

\begin{equation}
    V_\theta=\kappa(1+\cos{\theta})
    \label{bending}
\end{equation}
with $\theta$ being the angle between the three involved consecutive atoms and $\kappa$ being the coefficient of stiffness. 
A cosine type bending potential like Eq.~\eqref{bending} originates from the well-known Kratky-Porod model \cite{Kratky_49,DoiEdwards,RubinsteinColby} and is a common choice for modelling semiflexibility in polymers \cite{Auhl2003jcp}.
Note that even though our short flexible chains are essentially unentangled, Ref.~\cite{Faller_CPC_2001} suggests for a very similar model that the entanglement length drops steeply with increasing stiffness for semiflexible chains, implying that chains of length $N=15$ are already entangled for our intermediate range of stiffnesses $(\kappa\approx5)$.

Non-equilibrium molecular dynamics simulations of a sheared melt at density $\rho=0.8$ were performed using the LAMMPS simulation package \cite{Plimpton1995}. System sizes were set to $(15\sigma)^3$ if not mentioned otherwise. Shear along the x-direction was introduced by superimposing a velocity gradient on thermal velocities using the SLLOD equations \cite{Evans1984,Ladd1984,Tuckerman1997,Evans2008} and coupling the latter to the Nose-Hoover thermostat \cite{Evans1985,Tuckerman1997}. 
Temperature T=1 was maintained throughout our simulations, and the Velocity Verlet algorithm was used to integrate the equations of motion. Note, that LAMMPS implements a non-orthogonal simulation box with periodic boundary conditions that deforms continuously in accordance with the applied shear rate \cite{Evans1979,Hansen1994} - an approach that has been shown to be equivalent to the commonly used Lees-Edwards boundary conditions \cite{Evans2008,Todd2017}.

Shear viscosity $\eta(\dot{\gamma})$ was calculated using the relation 

\begin{equation}
\eta=\frac{\sigma_{xy}}{\dot{\gamma}},
\end{equation} 
where $\dot{\gamma}$ is the applied shear rate and $\sigma_{xy}$ is a non-diagonal component of the stress tensor as determined by the Irving-Kirkwood formula \cite{Irving1950,allen-tildesley-87}:

\begin{equation}
  \sigma_{xy}=-\frac{1}{V}\left[\sum_{i}^{N}\left(m_{i}v_{i,x}v_{i,y}\right)+ \sum_{i}^{N}\sum_{j>i}^{N}\left(r_{ij,x}f_{ij,y}  \right)    \right].
\label{vis}
\end{equation}
Here, $m_{i}$ is the mass of the $i^{th}$ particle, $\bf{v_{i}}$ the peculiar velocity of the $i^{th}$ particle, and $\bf{r_{ij}}$ and  $\bf{f_{ij}}$ are the distance and force vectors between the $i^{th}$ and the $j^{th}$ particle, respectively. For comparison, we have also calculated shear viscosity 
via the Green-Kubo relation:

\begin{equation}
\eta_{GK}=\frac{V}{k_BT}\int_{0}^{\infty}\left\langle\sigma_{xy}(t)\sigma_{xy}(0)\right\rangle dt.
\label{GK}
\end{equation}
where $V$ is the volume of the system and $k_B$ is the Boltzmann constant.
$\eta_{GK}$ is measured under equilibrium conditions ($\dot{\gamma}=0$) and serves as a reference value for $\eta(\dot{\gamma}\rightarrow{0})$.
Note that forces arising from the thermostat and its coupling to the SLLOD conditions are not explicitly considered in Eq.~\ref{vis} and may potentially result in a small systematic error. For detailed discussion of this effect in the context of dissipative thermostats the reader is referred to Ref.~\cite{Jung_2016}.

In addition, we have carried out Brownian dynamics simulations of 
single chains (same potentials as above) in an external shear flow profile. The equation of motion of monomers $i$ is given by

\begin{equation}
    \dot{\mathbf{r}_i} = \frac{1}{\zeta} \mathbf{f}_i + \dot{\gamma} z_i + \xi_i(t),
\end{equation}
where $\zeta$ is the monomeric friction, $\mathbf{f}_i$ the intermolecular force acting on $i$, and $\xi_{i,\alpha}(t)$ an uncorrelated Gaussian white noise with mean zero obeying the fluctuation-dissipation theorem, i.e.,
$\langle \xi_{i \alpha}(t) \xi_{i \beta}(t') \rangle = (2 k_B T /\zeta) \:  \delta_{ij} \delta_{\alpha \beta} \delta(t-t')$. In the simulations, we used an Euler forward algorithm with a time step $\Delta t = 10^{-4} \zeta \sigma^2/\epsilon$. The time scales of the two models were adjusted by mapping the Rouse times of fully flexible chains ($\kappa = 0$) at equilibrium ($\dot{\gamma} = 0$). To determine the Rouse time, we determined the autocorrelation function of the squared end-to-end distance,

\begin{equation}
    C_{R_{ee}}(t) = \frac{ \langle R_{ee}^2(t) R_{ee}^2(0) \rangle - \langle R_{ee}^2 \rangle^2 }{\langle R_{ee}^4 \rangle - \langle R_{ee}^2 \rangle^2} .
\end{equation}
For ideal Rouse chains, it can be calculated analytically, giving

\begin{equation}
    C_{R_{ee}}(t) = \Big( \frac{8}{\pi^2} \sum_{p \mbox{\tiny odd}} \frac{1}{p^2} \: \mathrm{e}^{- p^2 (t/\tau_R)^2} \Big)^2,
\end{equation}
where $p$ sums over the Rouse modes of the chain. At late times, the behavior is dominated by the first Rouse mode with $p=1$. We thus fitted the late time behavior of $\sqrt{C_{R_{ee}}(t)}$ to the function $A \mathrm{e}^{-t/\tau_R}$
for  chains of length $N=15$ in a melt and for the corresponding single Brownian chains. The fit parameters for the prefactor $A$  were in rough agreement with the theoretical value $8/\pi^2 = 0.81$ in both cases ($A = 0.87$ for melt chains, and $A=0.90$ for single oligomers). The fitted Rouse time of melt chains was $\tau_R = (85.4 \pm 0.1) \sqrt{m \sigma^2/\epsilon}$, and that of single oligomers was $\tau_R = (8.45 \pm 0.02) \zeta \sigma^2/\epsilon$. Hence the time scales match when choosing $\zeta = 10.1 \sqrt{m \epsilon/\sigma^2}$.
We have also carried out a more intricate mapping (discussed in Appendix A), which matches Rouse times for each value of $\kappa$, but does not change our results qualitatively.
\section{Shear-thinning in Oligomer Melts - a Molecular Analysis}
\label{sec_3}

In the following section we would like to investigate and review the molecular foundation of shear-thinning in low molecular weight polymer melts. We will show and highlight that macroscopic flow properties of polymers are governed by an intricate interplay of stretching, alignment and tumbling of individual molecules as well as collective modes at the molecular level.

\begin{figure}[ht]
\includegraphics[width=2.6in]{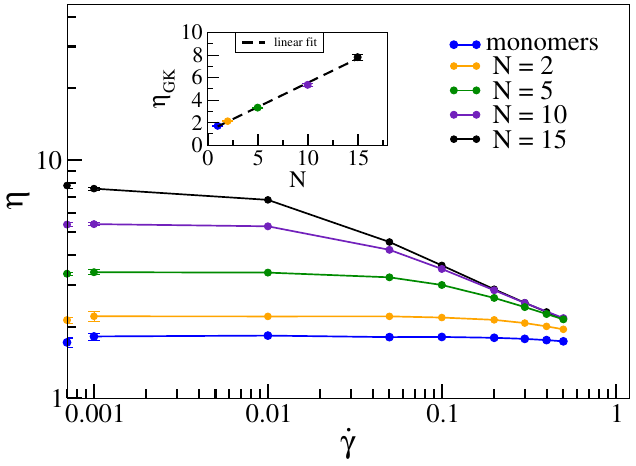}
\includegraphics[width=2.7in]{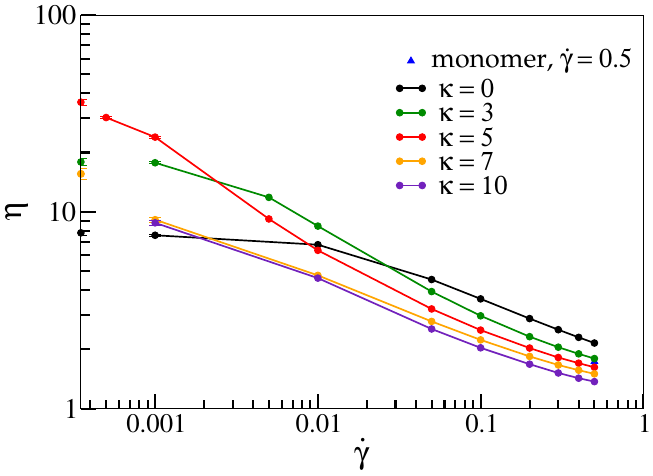}
\caption{ \textbf{(a)} Viscosity $\eta$ as function of shear rate $\dot{\gamma}$ for a melt of flexible oligomer chains ($\kappa=0$) with N= 1, 2, 5, 10 and 15 beads per chain. Corresponding shear viscosities according to the Green-Kubo relation $\eta_{GK}$ are shown on the y-axis and displayed as a function of $N$ in the inset. Density $\rho=0.8$ and box dimensions are $10\times10\times10\sigma^{3}$ for N= 1, 2, 5 and 10 and  $15\times15\times15\sigma^{3}$ for N=15.
\textbf{(b)} $\eta(\dot{\gamma})$ for a melt with N=15 and $\rho=0.8$ and varying stiffnesses. $\eta_{GK}$ for $\kappa=0,3,5,7$ and $10$ are displayed on the y-axis. The viscosity for monomers at $\dot{\gamma}$ = 0.5 (blue triangle) is also shown for reference. If not displayed explicitly, errors are smaller than symbol sizes. All lines serve as guides to the eye.
\label{fig1}
}
\end{figure}

Fig.~\ref{fig1}a displays viscosity $\eta$ as a function of shear rate $\dot{\gamma}$ for flexible oligomers with different chain lengths $N$. Flexible oligomers exhibit shear-thinning \cite{Yamamoto_2002}, i.e., decrease of viscosity with increasing shear rate. For consistency, we also compare viscosities as derived from Eq.~\eqref{vis} with those obtained from the Green-Kubo relation, Eq.~\eqref{GK} (points on the y-axis). The latter agree with the values for ${\dot{\gamma}}=0.001$ within error bars. 
The overall shape of $\eta(\dot{\gamma})$, namely a plateau at low shear rates followed by a shear thinning regime which becomes more pronounced with increasing molecular weight, has also been observed for various polymers experimentally \cite{Kim_Macro_1993, Heberer_JMatSci_1994, Bair_2002}.The inset shows that $\eta_{GK}$ increases linearly with $N$ for small chain lengths in agreement with simulations from \cite{Kroger_1993}.
%

In Fig.~\ref{fig1}b we investigate the dependence of viscosity on shear rate as a function of stiffness for an oligomer melt with chain length $N=15$. While for large shear rates, viscosity decreases with increasing stiffness and intriguingly even drops below the value determined for monomers (for $\kappa>3$), $\eta(\kappa)$ becomes non-monotonic for low shear rates. For ${\dot{\gamma}}=0.001$, flexible chains with $\kappa=0$ have the lowest viscosity, while viscosity increases for semiflexible chains and drops down again for rigid chains, an effect described in \cite{Xu_2017} for a similar model. A similar non-monotonic behaviour is exhibited by $\eta_{GK}$ (shown on the margins of Fig.\ref{fig1}b as a function of $\kappa$.) While the rise of viscosity can already be explained with the emergence of entanglements for intermediate stiffnesses, at the end of this section we will associate the following decline with a collective alignment of chains (and associated disentanglements) which are amplified by shear (Fig.~\ref{fig4}).

\begin{figure}[ht]
\includegraphics[width=2.6in]{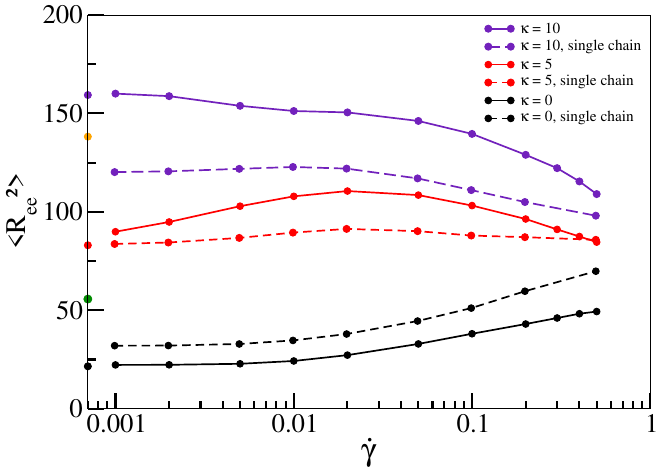}
\includegraphics[width=2.6in]{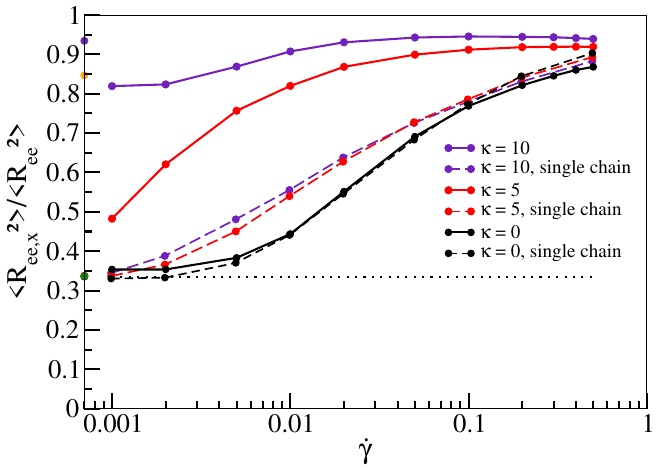}
\caption{\textbf{(a)} $\langle R_{ee}^2 \rangle$ as a function of shear rate $\dot{\gamma}$ for stiffnesses $\kappa=0$ and $\kappa=5$ at density $\rho=0.8$. \textbf{(b)} Ratio of the x-component $\langle R_{ee,x}^2 \rangle$ and $\langle R_{ee}^2\rangle$ as a function of $\dot{\gamma}$ for $\kappa=0$ and $\kappa=5$. The dotted line at the ratio
 of 1/3 marks the value for an unsheared melt. Results for a single chain in shear flow are shown as dashed lines (with points) in both figures. Values on the y-axis (in \textbf{a} and \textbf{b}, color scheme like in Fig.~\ref{fig1}b) correspond to equilibrium simulations without shear. As there is no preferred orientation in the bulk, the value for the ratio refers to the largest component. For $\kappa\leq5$ there is no preferred orientation in the bulk.
 All lines serve as guides to the eye.}
\label{fig2}
\end{figure}

In Fig.~\ref{fig2}a we quantify the stretching of individual chains with shear. The size of a flexible chains as measured by the mean square end-to-end distance $\langle R_{ee}^2\rangle$ increases continuously as a function of shear rate. For $\kappa=5$ the average size only increases slightly towards moderate shear rates before decreasing again at high rates similar to \cite{Kong2019}, ruling out stretching as a main driving force for shear-thinning in this regime for melts of semiflexible chains. 
Fig.~\ref{fig2}b visualizes the alignment of chains along the shear direction by plotting the ratio of the x-component to the total $\langle R_{ee}^{2}\rangle $. Without shear each component contributes equally yielding a ratio of 1/3 (dotted line in Fig.~\ref{fig2}b). While this holds for flexible chains at low shear rates, deviations become more pronounced for shear rates exceeding $0.01$. This is also roughly the rate at which noticeable deviations from $\eta_{GK}$ start to occur in Fig.~\ref{fig1}a and shear-thinning sets in. This behaviour becomes even more pronounced for semiflexible chains. At $\dot{\gamma}=0.001$ chains are already partially aligned and the viscosity in Fig.~\ref{fig1}b already deviates significantly from the value obtained from the Green-Kubo relation. Shear-thinning sets in at even lower shear rates and is reinforced with progressive alignment of chains. This relation provides clear indication that chain alignment is strongly correlated with the occurrence of shear-thinning in agreement with previous observations \cite{Xu_2017,Kong2019}. 
Note that for $\kappa>5$ chains are already stretched and aligned in equilibrium simulations without shear (values on y-axis of Fig.~\ref{fig2}~a,b) indicating the emergence of nematic behaviour in agreement with previous observations in a similar model \cite{Faller_Macro_2000,Faller_CPC_2001}. For $\kappa=10$ there is even an initial drop from the bulk ratio once shear sets in.
Interestingly, stretching and alignment of chains can already be observed qualitatively in simulations of single chains in shear flow (dashed lines in Figs.~\ref{fig2}a,b) 
indicating that these phenomena 
should in principle be observable for melts of all densities. 
However, while the alignment of flexible chains is well-reproduced, the increase is less pronounced for higher stiffnesses, indicating that collective alignment contributions due to stiffness are not captured by single chain simulations.
%
In the following we turn to movement modes of individual chains. 
\begin{figure}[ht]
   \includegraphics[width=2.8in]{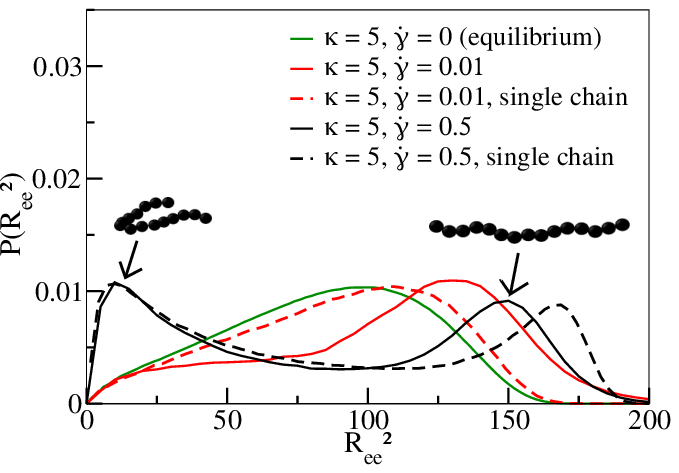}
   \includegraphics[width=2.8in]{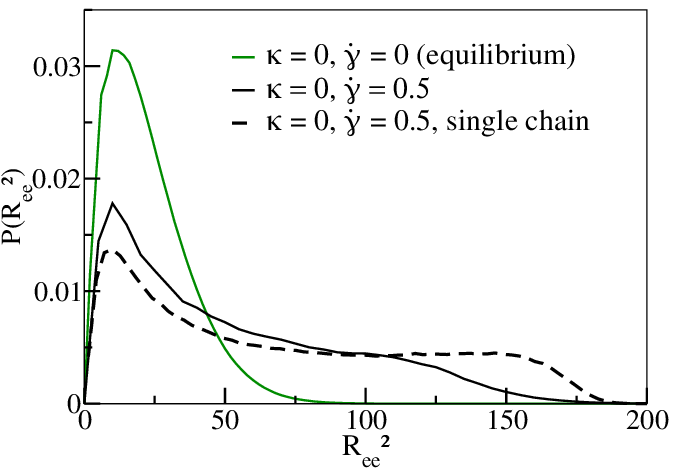}
   \caption{\textbf{(a)} Probability distributions P($R_{ee}^{2}$) for stiffness $\kappa$ = 5 at shear rates $\dot{\gamma}=0, 0.01$ and $0.5$. The two peaks of the distribution for $\dot{\gamma}=0.5$ correspond to U-shaped configurations and stretched configuration of individual oligomers, respectively, as indicated by typical snapshots. \textbf{(b)} P($R_{ee}^{2}$) for stiffness $\kappa$ = 0 and shear rate $\dot{\gamma} = 0$ and $0.5$. Results for a single chain in shear flow are shown as dashed lines in both figures.}
   \label{fig3}
\end{figure}

Fig.~\ref{fig3} shows the distribution of $R_{ee}^{2}$ of individual chains in melts. 
While equilibrium simulations ($\dot{\gamma}=0$) have a broad distribution of end-to-end distances for $\kappa=5$ (solid green line in Fig.~\ref{fig3}a), conformations develop a preference for stretched chains at moderate shear rates ($\dot{\gamma}=0.01$, solid red line). At about this rate, $\langle R_{ee}^{2}\rangle$ displays a maximum in Fig.~\ref{fig2}a. For the highest shear rate $\dot{\gamma}=0.5$ (solid black line), U-shaped tumbling conformations coexist with stretched conformations explaining the decrease of $\langle R_{ee}^{2}\rangle$ in Fig.~\ref{fig2}a for $\dot{\gamma}>0.02$ \cite{Kong2019}. For flexible chains compact conformations dominate the behaviour at small and large shear rates (Fig.~\ref{fig3}b) as already noted in \cite{Yamamoto_2002}, even though the latter also exhibit some degree of stretched conformations. This also explains why $\langle R_{ee}^{2}\rangle$ is significantly smaller in comparison to semiflexible chains (Fig.~\ref{fig2}a). It is worth noting that the occurrence of compact conformations does not impede the continuous alignment of chains along the shear direction with increasing shear rate as demonstrated in Fig.~\ref{fig2}b. The intricate interplay between stretching and tumbling as a function of shear rate and stiffness can also be observed for our single chain simulations (as noted for flexible chains already in \cite{Yamamoto_2002} and studied in \cite{Celani2005}), indicating that these movement modes are not a collective phenomenon and should occur in melts of all densities \cite{Xu2020}.

\begin{figure}[ht]
   \includegraphics[width=2.6in]{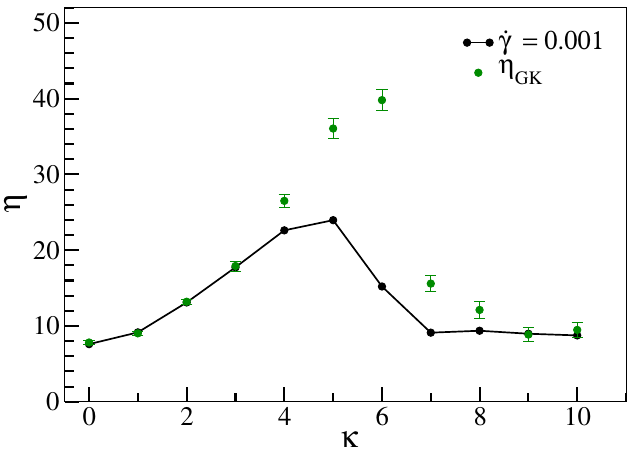}
   \includegraphics[width=2.7in]{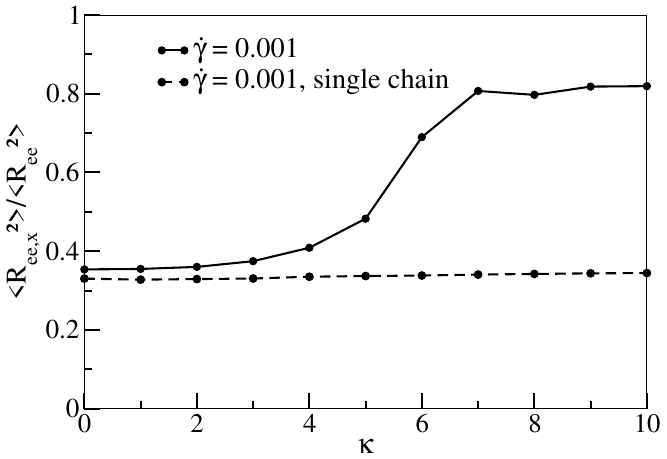}
   \caption{\textbf{(a)} Viscosity $\eta$ as a function of stiffness $\kappa$ for shear rate $\dot{\gamma}$ = 0.001. Shear viscosity at zero shear rate $\eta_{GK}$  are shown as green dots. \textbf{(b)} $\langle R_{ee,x}^2\rangle / \langle R_{ee}^2\rangle$ as a function of $\kappa$ at $\dot{\gamma} = 0.001$ for melt and single chain simulations (dashed lines). All lines serve as guides to the eye.}
   \label{fig4}
\end{figure}

In Fig.~\ref{fig4}, we finally investigate the non-monotonous behaviour of the Green-Kubo viscosity $\eta_{GK}$ and viscosity at low shear rates ($\dot{\gamma} = 0.001$) as functions of $\kappa$. $\eta_{GK}$ increases with increasing chain stiffness, reaches a maximum at about $\kappa=6$ and undergoes a sharp decrease after that. Also, $\eta(\dot{\gamma}=0.001)$ increases up to $\kappa=5$ and decreases subsequently. Intriguingly, $\eta_{GK}$ matches with $\eta(\dot{\gamma}=0.001)$ up to $\kappa=3$, but, between $\kappa=4$ and $\kappa=6$, $\eta_{GK}$ is significantly larger than $\eta(\dot{\gamma}=0.001)$. As already pointed out before, Ref.~\cite{Faller_CPC_2001} estimates that the entanglement length decreases with increasing  persistence length, $l_{p}$, to a point at which the entanglement length becomes smaller than the chain size. It estimates that at $l_{p}=1.5, 3, 5$, the entanglement lengths are approximately equal to 15, 8 and 6, respectively. Entanglement length decreases with increasing $l_{p}$. Ref.~\cite{Faller1999} estimates that the numerical values of $l_{p}$ are quite close to the numerical values of $\kappa$. For example, $\kappa=3$ and $5$ corresponds to $l_{p}\approx$  2.5 and $\approx5$, respectively. So, chains become entangled and this effect increases with increasing $\kappa$. It should be noted, however, that for both Ref.~\cite{Faller_CPC_2001} and Ref.~\cite{Faller1999}, number density was 0.85, which is a bit higher than the number density of our system (0.8). Bond and angular potentials forms also differ slightly in \cite{Faller_CPC_2001}. Unaligned and entangled chains impede collective motion under equilibrium conditions, and as a result $\eta_{GK}(\kappa)$ increases up to $\kappa=6$. Following $\kappa=6$, however, there is a sharp drop in $\eta_{GK}$. This is consistent with our observation from Fig.~\ref{fig2}b that for $\kappa=7$ and $\kappa=10$, chains are already stretched and aligned under equilibrium conditions indicating that our system indeed undergoes an isotropic-nematic transition following $\kappa=6$. Entanglements decrease as chains align and conformations become more susceptible to the applied shear, resulting in a decrease in $\eta_{GK}$. $\eta(\dot{\gamma}=0.001)$ as a function of $\kappa$ also exhibits a similar non-monotonous behavior. While the initial increase can be also attributed to progressive entanglements, for $\kappa=4$, and $\kappa=5$, the applied shear already begins to align the chains towards the shear direction, counteracting entanglement effects, as is evident in Fig.~\ref{fig4}b. As a result $\eta(\dot{\gamma}=0.001)$ for $\kappa=4$ and $\kappa=5$ are lower than the corresponding $\eta_{GK}$ values.  For $\kappa>5$, chains are more strongly aligned along the shear direction, thus $\eta(\dot{\gamma}=0.001)$ as a function of $\kappa$ decreases beyond $\kappa=5$ \cite{Xu_2017}. This behaviour is, in contrast to stretching and alignment with increasing shear rate, a collective phenomenon which cannot be observed in corresponding simulations of single chains and should therefore vanish gradually at smaller densities.


\section{Hybrid multiscale method}
\label{sec_4}

After studying the molecular origin of non-Newtonian behaviour of short polymer melts, we turn now our attention to  macroscopic simulations combining them with the results of Molecular Dynamics.

The motion of an incompressible fluid flow at the macroscopic level is governed by the continuity and the momentum equations
\label{sec:cfd}
\begin{subequations}
\begin{alignat}{3}
\label{eq:Continuity}
&\nabla \cdot \vec{u}=0,&\quad &\textrm{in} \; \Omega \times [0, T]  \\ 
\label{eq:Momentum}
&\frac{\partial \vec{u}}{\partial t}+ \vec{u}\cdot\nabla\vec{u}= \nabla \cdot
\grvec{\sigma}+\vec{g},\quad& &\textrm{in} \; \Omega \times [0, T]  \\  
&\vec{u}=\vec{u}_D, & &\textrm{on} \; \partial\Omega_D\\ 
&\grvec{\sigma} \cdot \vec{n} = 0, & &\textrm{on} \; \partial\Omega_N\\
&\vec{u}(t=0)=\vec{u}^{(0)} & &\textrm{in} \; \Omega,  \label{eq:initial}
\end{alignat}
\end{subequations}
\noindent
where $\vec{u}$ is the velocity vector, $\grvec{\sigma}$ the Cauchy stress tensor, and $\vec{g}$ an external body force.
Boundary of the computational domain $\Omega$ consists of the Dirichlet, Neumann and periodic boundary, i.e.~
$\partial \Omega = \partial\Omega_D \cup \partial\Omega_N \cup \partial\Omega_P$.

The Cauchy stress tensor can be split into two parts $\grvec{\sigma}=-p \textbf{I}+\grvec{\tau}$ with $p$ being an isotropic hydrostatic pressure  and $\grvec{\tau}$ an viscous stress tensor.
For the Navier-Stokes equations we have $\grvec{\tau}=\eta(\nabla \vec{u}+\nabla \vec{u}^T)$ with $\eta $ being a
constant viscosity. This relation is more complex when non-Newtonian polymer fluids are considered. 

In this work we apply the \emph{ hybrid multiscale method} that couples the Molecular Dynamics simulations with the macrocopic model \eqref{eq:Continuity}-\eqref{eq:initial}. As explained in Section~\ref{sec_2} the macroscopic stress tensor can be derived from the Irving-Kirkwood formula \eqref{vis}.

\noindent { Our extensive Molecular Dynamics simulations imply that the stress tensor can be actually expressed in the following simple way}

\begin{equation}
\label{eq:tau}
\grvec{\tau}= \eta(\dot{\gamma}) (\nabla \vec{u}+\nabla \vec{u}^T).
\end{equation}

\noindent
{ Finally, the viscosity--shear rate dependence leads  to a well-known Carreau-Yasuda rheological model} \cite{andrade2007}

\begin{equation}
\label{eq:carreau}
\eta(\dot{\gamma}) = \eta_\infty + (\eta_0-\eta_\infty) (1+(a_2\dot{\gamma})^{a_3})^{a_1}
\end{equation}
\noindent
with the following coefficients:
for flexible polymers (stiffness $\kappa=0$)
$\eta_0=7.76$,
$\eta_\infty = 1.08441$,
$a_1 = -0.425387$,
$a_2 = 54.3905$,
$a_3 = 1.28991$;
and for semiflexible  polymers with $\kappa$=5
$\eta_0=36.052$,
$\eta_\infty = 1.09319$,
$a_1 = -0.214969$,
$a_2 = 2143.96$,
$a_3 = 2.78713$.

\noindent { We note by passing that  for a particular situation considered in this paper our hybrid multiscale method can be seen as a parameter passing sequential coupling multiscale method, see, e.g.~\cite{E2007} for a detailed description of the concurrent and sequential coupling strategies.} 

\begin{figure}[h!]
\centering	\includegraphics[width=3.0in, clip]{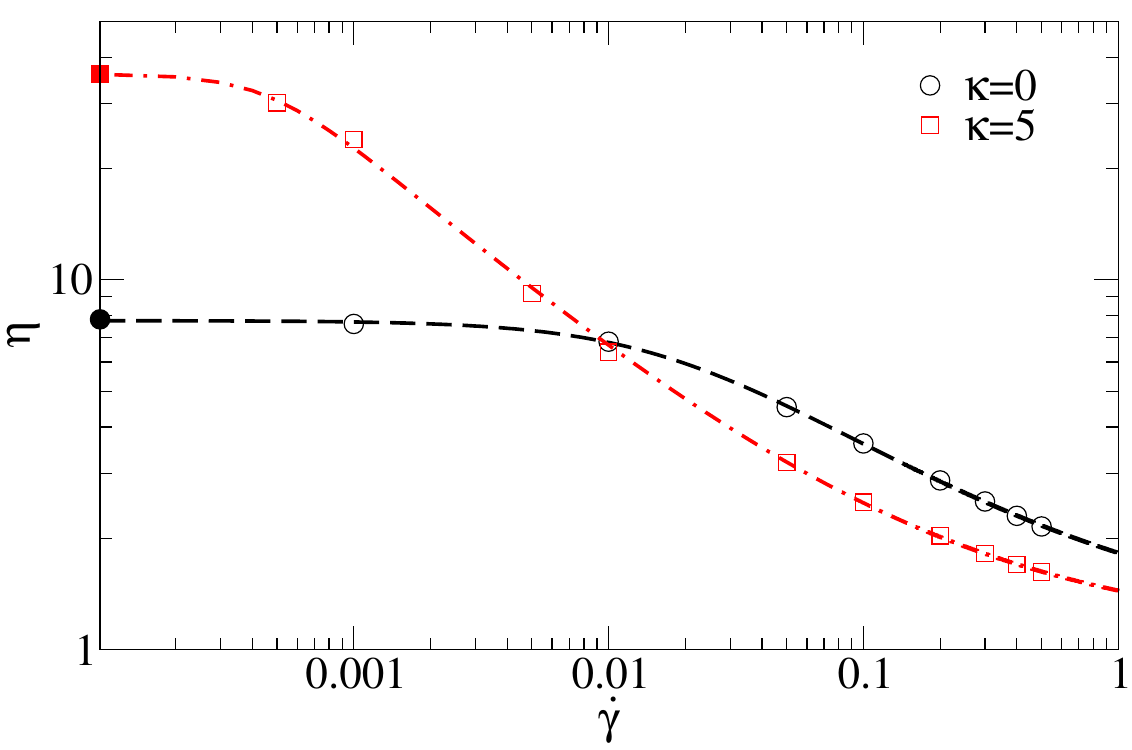}
\caption{Shear viscosity of flexible oligomer chains and oligomer chains with stiffness $\kappa=5$ as obtained by non-equilibrium Molecular Dynamics (open symbols) and fitting by the Carreau-Yasuda rheological fluid model Eq.~\eqref{eq:carreau} (dashed curves, compare with Fig.~\ref{fig1}b). 
Bold symbols on the y-axis represent viscosity values at $\dot{\gamma}=0$ obtained from equilibrium Molecular Dynamics simulations.}
\label{fig:carreau}
\end{figure}

\noindent
The oligomer chain length in both cases is $N=15$. Fig.~\ref{fig:carreau} compares the MD data and the fitting with Carreau-Yasuda model.

Our next goal is to calculate the shear-dependent viscosity $\eta(\dot{\gamma}).$  
In what follows we consider for simplicity  the situation of two-dimensional shear flows and use the notation $\vec{u} = (u,v)$.
Applying \eqref{eq:carreau}  we need the  value of the shear rate $\dot\gamma$ of the polymer flow. 
It can be obtained from the strain-rate tensor

\begin{equation}
\mS = \frac{\nabla \matr{u}+ \nabla \matr{u}^T} {2} =
\begin{pmatrix}
\frac{\partial u}{\partial x} & \frac{ \left(\frac{\partial u}{\partial y}+\frac{\partial v}{\partial x}\right) }{2} \\
\frac{ \left(\frac{\partial u}{\partial y}+\frac{\partial v}{\partial x}\right) }{2} & \frac{\partial v}{\partial y}
\end{pmatrix}
\label{eq:mS2d}
\end{equation}

\noindent
by rotating it with respect to the streamlines to the anti-diagonal matrix

\begin{equation}
\mS'=\mTheta\mS\mTheta^T=
\begin{pmatrix}
0 & \dot\gamma/2 \\
\dot\gamma/2 & 0
\end{pmatrix}\
\label{eq:mSprim}
\end{equation}

\noindent
by an angle $\theta$, which for incompressible flows is $\theta=\frac{1}{2}\tan^{-1}(-\frac{S_{xx}}{S_{xy}})$. Here $S_{ij}$ are components of the strain-rate tensor $\mS$ and $\mTheta$ the rotation matrix \cite{Todd1998, yasuda2008model}.
The new strain-rate tensor $\mS'$ corresponds to a pure-shear deformation (i.e., in absence of normal stresses).
The shear rate can therefore be calculated from the components of the original strain-rate tensor $\mS$ and the angle $\theta$

\begin{equation}
\dot\gamma/2= S_{xy}\cos(2\theta) -S_{xx}\sin(2\theta).
\label{eq:dotgamma}
\end{equation}

\noindent
The stress tensor in the shear flow can be calculated in the normal-stress free basis and transformed back to the original basis according to \eqref{eq:tau}

\begin{equation}
\grvec{\tau}=\mTheta^T \left(2\eta(\dot\gamma) \mS'\right) \mTheta.
\label{eq:tau2}
\end{equation}

As shown in \eqref{eq:tau2} we consider in the present work shear dependent flows, where the stress tensor or more precisely the viscosity are nonlinear functions of the local shear rate. In order to consider more general flow conditions such as the extensional flow, rigid rotation and mixed flows one needs to take into account not only the shear rate dependence but  a complete decomposition of a three-dimensional symmetric tensor (stress tensor) into a six dimensional basis. In such a way not only the viscosity but also additional response coefficients will be computed from  microscopic simulations in order to determine the local stress tensor.
We refer a reader to our recent work \cite{giulio} where  complex flows in general geometeries were studied, see also \cite{GS_18}.

We proceed by describing a numerical method applied to \eqref{eq:Continuity}-\eqref{eq:initial}. For \emph{ time integration} we apply the implicit BDF2 scheme which leads to the following system

\begin{subequations}
\begin{alignat}{5}
\label{NonlinearStep-Euler}
&\mbox{(1)} \quad 
&&\frac 3 2 \vec{u}^{(n+1)} + \Delta t 
\big(\vec{u}^{(n+1)}\cdot\nabla\vec{u}^{(n+1)} + \nabla p^{(n+1)} - \nabla \cdot \grvec{\tau}^{(n+1)}-\vec{g}^{(n+1)}\big) 
\\  \nonumber
&\phantom{mm}&&=2\vec{u}^{(n)} - \frac 1 2 \vec{u}^{(n-1)} \quad \textrm{in} \; \Omega&&\\
&\mbox{(2)} \quad &&
\nabla \cdot \vec{u}^{(n+1)} = 0  \quad \qquad \textrm{in} \; \Omega.&&
\end{alignat}
\end{subequations}

Discretization in space is realized by the discontinuous Galerkin (dG) method \cite{emamy-tuprints3471, Emamy2017285, Stalter2018}.
Domain $\Omega$ is discretized by a quadrilateral mesh $\mathcal{T}_h$ with a meshsize $h$.  Mesh faces $\mathcal{F}_h$ can be either the inner interfaces between adjacent elements $\mathcal{F}_h^{i}$ or boundary faces $\mathcal{F}_h^b$. The face normal $\vec{n}_\mathcal{F}$ points from an arbitrarily chosen (but fixed) element $T_1$ towards $T_2$.  The face normal points outward of the domain $\Omega$ on boundary faces.

We consider  broken Sobolev spaces $V_h,\, Q_h$ that consist of piecewise quadratic and piecewise linear polynomials, respectively,

$$
V_h:= \{v \in L^2(\Omega)| \,\mbox{ for all } T \in \mathcal{T}_h, v|_T \mbox{ is a quadratic polynomial} \}
$$
$$
Q_h:= \{q \in L^2(\Omega)| \,\mbox{ for all } T \in \mathcal{T}_h, q|_T \mbox{ is a linear polynomial} \}.
$$

The usual average and jump operators of a scalar-valued function $f_h$ on interfaces between adjacent elements $T_1$ and $T_2$ are defined as

$$
\llrrbrace{f_h}=\frac{1}{2}(f|_{T_1}+f|_{T_2}),\qquad 
\llrrbracket{f_h}=f|_{T_1}-f|_{T_2}.
$$
Vector-valued functions are treated componentwisely. For boundary faces, we set $\llrrbrace{f_h}=\llrrbracket{f_h}=f|_{T}$, when not mentioned otherwise. 

\medskip

\noindent \textbf{\emph{Discontinuous Galerkin (dG) method}} for an oligomer fluid flow  \eqref{eq:Continuity}-\eqref{eq:initial} is formulated as follows.
 Given the initial data $\vec{u}_h^{(0)} \in V_h$ we look for a sequence of numerical solutions $\vec{u}_h^{(n+1)} \in V_h,$ $p_h^{(n+1)} \in Q_h $ for $n= 1,\dots, N_T-1$ such that
 
\begin{align}
& \left(\frac 3 2\vec{u}_h^{(n+1)}, \grvec{\varphi}_h\right) +  \Delta t \, \big(\,c(\vec{u}_h^{(n+1)},\vec{u}_h^{(n+1)},\grvec{\varphi}_h)
- b(p_h^{(n+1)}, \grvec{\varphi}_h )
+ a( \eta(\dot{\gamma_h}^{(n+1)}), \vec{u}_h^{(n+1)},\grvec{\varphi}_h) \big)
\nonumber \\ \nonumber
&\phantom{mmmmmm}= \Delta t \, (\vec{g}_h^{(n+1)}, \grvec{\varphi}_h) 
+  \left(2 \vec{u}_h^{(n)} - \frac 1 2 \vec{u}_h^{(n-1)}, \grvec{\varphi}_h\right) \nonumber \\ 
&  b(q_h, \vec{u}^{(n+1)}_h ) = 0  
\qquad \qquad \mbox{  for any } \ \grvec{\varphi}_h \in V_h, \, q_h \in Q_h.
\label{dG}
\end{align} 
The numerical shear rate $\dot{\gamma_h}^{(n+1)}$ is computed locally, i.e. in each quadrature point
from the broken gradients $\nabla \vec{u}_h^{(n+1)}$.
Hereby we compute the first approximation $\vec{u}_h^{(1)}$, e.g., by the Euler implicit method.
The $L^2(\Omega)$  scalar product is denoted by  $(\cdot\, , \cdot).$  
In what follows we define the discrete forms 
$a$, $b$, and $c.$

\noindent The {\sl convective term} is rewritten  in the  conservative form and the interface integrals are approximated  by means of the Lax-Friedrich numerical flux

$$
c(\vec{u}_h,\vec{w}_h,\grvec{\varphi}_{h})=-\int_{\Omega} (\vec{u}_h \otimes \vec{w}_h) : \nabla\grvec{\varphi}_{h}+\int_{\mathcal{F}_h} \big(\llrrbrace{\vec{u}_h \otimes \vec{w}_h}\cdot \vec{n}_\mathcal{F}+\frac{1}{2}\Lambda \llbracket \vec{u}_h\rrbracket \, \big)\cdot \llbracket\grvec{\varphi}_{h}\rrbracket,
$$
where $\Lambda=\max(\lambda|_{T_1},\lambda|_{T_2})$ and $\lambda$ is the absolute eigenvalue of the Jacobian matrix $ \big(\partial [(\vec{u} \otimes \vec{w}) \cdot \vec{n}_\mathcal{F}]/\partial \vec{u}\big)|_{\bar{\vec{u}},\bar{\vec{w}}}$. The average and jump operators on the Dirichlet boundaries are defined as

\begin{align*}
&\llrrbrace{\vec{u}_h \otimes \vec{w}_h}=\displaystyle\frac{1}{2}\big((\vec{u}_h \otimes \vec{w}_h)|_T+(\vec{u}_{D}\otimes \vec{w}_D)\big), \ \llbracket \vec{u}_h\rrbracket=(\vec{u}_h|_T-\vec{u}_{D}) 
  &\mathcal{F} \subset \partial\Omega_D\\
&\llrrbrace{\vec{u}_h \otimes \vec{w}_h}= (\vec{u}_h \otimes
\vec{w}_h)|_T , \ \llbracket \vec{u}_h\rrbracket=\vec{u}_h|_T 
  &\mathcal{F} \subset \partial\Omega_N\\
&\llbracket\grvec{\varphi}_{h}\rrbracket=\grvec{\varphi}_{h}|_T  
  &\mathcal{F} \subset \partial\Omega_D \cup \partial\Omega_N.
\end{align*}

\medskip

\noindent The {\sl divergence of the  velocity ${\vec{u}}_h$} 
as well as the {\sl discrete gradient of the pressure $p_h$} are both
approximated using the form

$$
b(r_h, \grvec{\phi}_h) =
- \int_{\Omega}{\grvec{\phi}}_h \cdot \nabla r_h
+ \int_{\mathcal{F}_h \setminus \partial \Omega_N} 
     \left\{\mkern-9.5mu\left\{{\grvec{\phi}}_h \right\}\mkern-9.5mu\right\} \cdot \vec{n}_\mathcal{F} \llbracket r_h \rrbracket.
$$

%
%

\noindent To discretize the viscous terms we employ an almost-standard symmetric
interior penalty method (SIP), first introduced in \cite{arnold_interior_1982} and 
extensively analyzed in \cite{arnold_unified_2002}

\begin{multline*}
a(\dot{\gamma}_h, \vec{u}_h,\grvec{\varphi}_h) =
 \int_\Omega   \eta(\dot{\gamma}_h) (\nabla \vec{u}_h  + \nabla \vec{u}_h^T ) : \nabla \grvec{\varphi}_h 
 + \int_{\mathcal{F}_h \setminus \partial \Omega_N }  \mu_P
     \eta(\dot{\gamma}_h)  \llrrbracket{ \vec{u}_h } \cdot \llrrbracket{ \grvec{\varphi}_h }
\\
- \int_{\mathcal{F}_h \setminus \partial \Omega_N }
        \left( \llrrbrace{\eta(\dot{\gamma}_h) ( \nabla \vec{u}_h + \nabla \vec{u}_h^T)} \vec{n}_\mathcal{F} \right) \cdot \llrrbracket{ \grvec{\varphi}_h }
      + \left( \llrrbrace{ \eta(\dot{\gamma}_h) ( \nabla \grvec{\varphi}_h + \nabla \grvec{\varphi}_h^T)} \vec{n}_\mathcal{F} \right) \cdot \llrrbracket{ \vec{u}_h }
 .
\label{eq:SIP-viscous}
\end{multline*}
%
On the Dirichlet boundaries $\mathcal{F} \subset \partial\Omega_D$ we set

$$
\llrrbrace{\ldots}=(\ldots)|_T, \
\llrrbracket{\vec{u}_{h}}=(\vec{u}_{h}|_T- \vec{u}_{D}), \ 
\llrrbracket{\grvec{\varphi}_{h}}=\grvec{\varphi}_{h}|_T.
$$
Coefficient 
$\mu_P$ is a penalty parameter that we choose for a quadraliteral mesh in the following way \cite{shahbaziexplicit2005}

\begin{align}
&\mu_P=\alpha_P c=\left \{
\begin{array}{l l}
\alpha_P \max(c|_{T_1},c|_{T_2}) &\mathcal{F} \in \mathcal{F}_h^{i} \nonumber \\
\alpha_P \, c|_{T}, &\mathcal{F} \in \mathcal{F}_h^{b} \nonumber
\end{array}\right.\\
&\displaystyle c|_{T}=3^2\textrm{ }
\frac{A(\partial T \backslash \mathcal{F}_h^b)/2+A(\partial T \cap \mathcal{F}_h^b)}{V(T)}, 
\nonumber
\end{align}
where $\alpha_P\ge 1$ is a user defined small coefficient. Area and volume are denoted by $A$ and $V$, respectively. 	
%
%
%
 We conclude this section by mentioning that the resulting nonlinear system is approximated by the Newton method imploying the Dogleg-globalization and a direct sparse solver for the linearized system, see~\cite{florian}.

\section{Two channel flows of a non-Newtonian oligomer fluid}
\label{sec_5}

In this section we illustrate the consequences of shear-thinning, whose microscopic origins have been discussed in previous sections, on two examples of macroscopic channel flows.
Numerical simulations have been realized within BoSSS code, for more details see also \cite{Emamy2017285, emamy-tuprints3471, emamy_2017, Kummer, florian}. 

In the first test the so-called Poiseuille flow~\cite{batchelor} is simulated.
Here an oligomer melt flows through a narrow channel of length $\ell=1$, driven by a pressure difference between the
outlet and inlet of the channel.
The intensity of the flow is controlled by the pressure parameter related to the external pressure gradient
$P_\textrm{x}=(P_\textrm{in} - P_\textrm{out})/\ell = -\partial p/\partial x$.  Since the viscosity 
is a function of the shear rate we compute the Reynolds
number using the averaged viscosity $\bar{\eta} = (\eta_0 + \eta_\infty)/2,$ i.e.
$$ 
Re = \dfrac{U \, L}{ \bar{\eta}},
$$
where $U$ is the characteristic velocity (maximal inflow velocity) and $L$  the characteristic length, i.e.~the channel diameter $L=1$. 
In order to take into account
also the effects of asymptotic viscosity values, 
we define 
$Re_0 =  U L/\eta_0$, $Re_\infty = U L/\eta_\infty$
and introduce them in Table~1.

\begin {table}[h]
\caption {Reynolds numbers} \label{tab:reynolds} 
\centering
\begin{tabular}{|*{5}{c|}}
\hline
\multicolumn{1}{|c|}{} & 
\multicolumn{4}{|c|}{$\kappa=0$} \\ \hline
\multicolumn{1}{|c|}{} & 
\multicolumn{1}{|c|}{} & 
\multicolumn{1}{|c|}{$\eta_0=7.76$} & 
\multicolumn{1}{|c|}{$\eta_\infty=1.084$} & 
\multicolumn{1}{|c|}{$\bar{\eta}=4.422$} \\ 
\hline\hline
\multicolumn{1}{|c|}{$P_\textrm{x}$} & 
\multicolumn{1}{|c|}{$U$} & 
\multicolumn{1}{|c|}{$Re_0$} & 
\multicolumn{1}{|c|}{$Re_\infty$} & 
\multicolumn{1}{|c|}{$Re$} \\ \hline 
0.02 & 3.25e-4 & 4.18e-5 & 2.99e-4 & 7.34e-5 \\ \hline
0.2  & 3.70e-3 & 4.76e-4 & 3.41e-3 & 8.36e-4 \\ \hline
1  & 3.29e-2 & 4.24e-3 & 3.04e-2 & 7.45e-3 \\ \hline
\hline
\multicolumn{1}{|c|}{} & 
\multicolumn{4}{|c|}{$\kappa=5$} \\ \hline
\multicolumn{1}{|c|}{} & 
\multicolumn{1}{|c|}{} & 
\multicolumn{1}{|c|}{$\eta_0=36.05$} & 
\multicolumn{1}{|c|}{$\eta_\infty=1.093$} & 
\multicolumn{1}{|c|}{$\bar{\eta}=18.57$} \\
\hline\hline
\multicolumn{1}{|c|}{$P_\textrm{x}$} & 
\multicolumn{1}{|c|}{$U$} & 
\multicolumn{1}{|c|}{$Re_0$} & 
\multicolumn{1}{|c|}{$Re_\infty$} & 
\multicolumn{1}{|c|}{$Re$} \\ \hline 
0.02 & 7.08e-5 & 1.96e-6 & 6.48e-5 & 3.81e-6 \\ \hline
0.2  & 3.69e-3 & 1.02e-4 & 3.38e-3 & 1.99e-4  \\ \hline
1  & 5.49e-2 & 1.52e-3 & 5.03e-2 & 2.96e-3 \\ \hline
\end{tabular}

\end{table}

\begin{figure}[htbp]
\centering
\includegraphics[width=3.5in, clip]{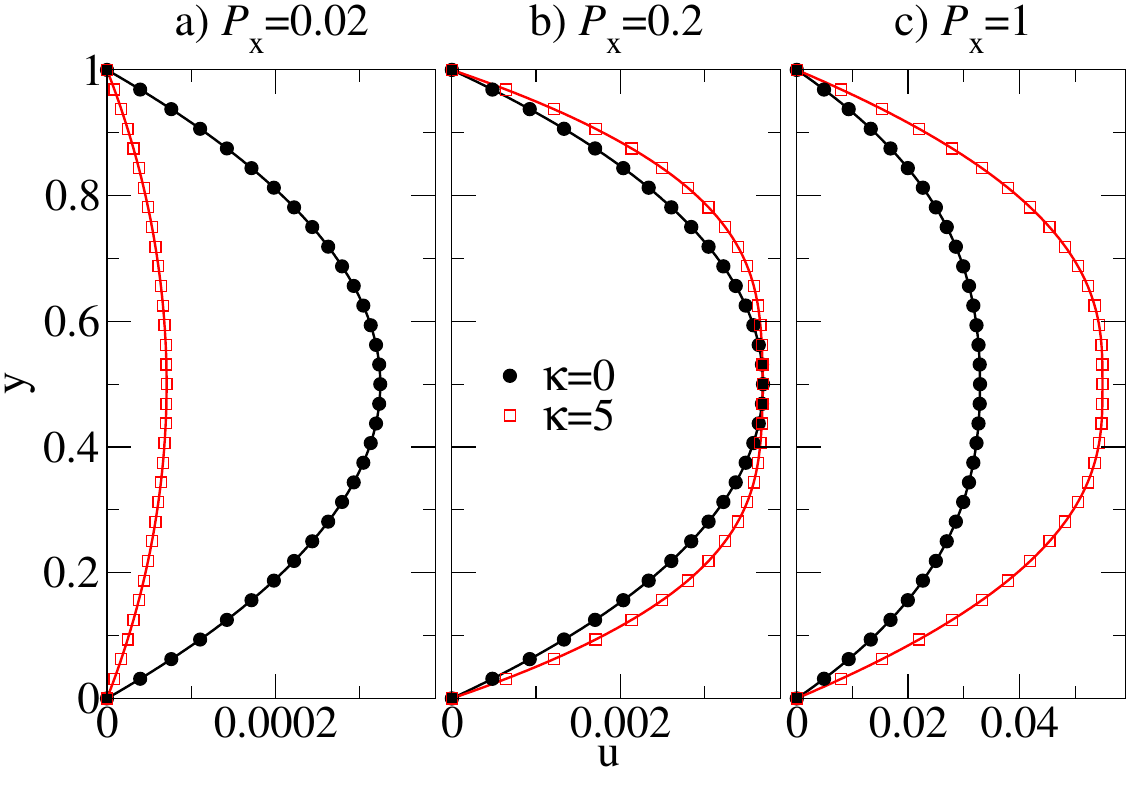}
\caption{
Steady-state velocity profiles of the pressure-driven channel flow of oligomer
melts consisting of either flexible or semiflexible chains with stiffness $\kappa=5$. Solutions are computed
by a hybrid MD-dG method (\ref{dG}) for various external pressure difference parameters $P_\textrm{x}$.
}
\label{fig:uprof}
\end{figure}

For the macroscopic model of our hybrid multiscale method we have used a grid with $128 \times 128$ mesh cells, i.e.~the mesh parameter $h=1/128.$ Numerical simulations on a coarser grid confirm that the steady state results presented here are independent on the mesh resolution.
Fig.~\ref{fig:uprof} shows the steady-state velocity profiles across the channel for flexible ($\kappa=0$)
and semiflexible ($\kappa=5$) polymer chains of the length $N=15$.
The external pressure drop is chosen at values of $P_\textrm{x}=\{0.02, 0.2, 1\}$ in order to
study the influence of the different regimes of the shear rate-dependent viscosity on an oligomer flow.
At low pressure drop ($P_\textrm{x}=0.02$ in Fig.~\ref{fig:uprof}a), the flow is very slow and the shear rates low, too.
The viscosity of both polymeric systems is at the Newtonian plateau (cf.~Fig.~\ref{fig:carreau}).
The  viscosity of flexible chains is lower than that of the semiflexible chains (ca. 8 vs. 36) and
therefore the velocity of the flexible chains is higher than that of the semiflexible chain melt.
At moderate pressure drop ($P_\textrm{x}=0.2$ in  Fig.~\ref{fig:uprof}b), the viscosity of the flexible chains is still nearly
a constant (the Newtonian regime) whereas the semiflexible chain melt is already in the shear-thinning regime.
By coincidence the amplitude of the two profiles at the center of the channel is approximately the same
and one can easily recognize that the shapes of the velocity profiles are very different:
the semiflexible chains exhibit a broader distribution (Fig.~\ref{fig:uprof}b).
Further increase of the pressure drop leads to an inverse situation:
the melt of semiflexible chain flows faster through the channel than the flexible chains melt
(cf.~$P_\textrm{x}=1$ in Fig.~\ref{fig:uprof}c).

\begin{figure}[htbp]
\centering
\includegraphics[width=2.5in, clip]{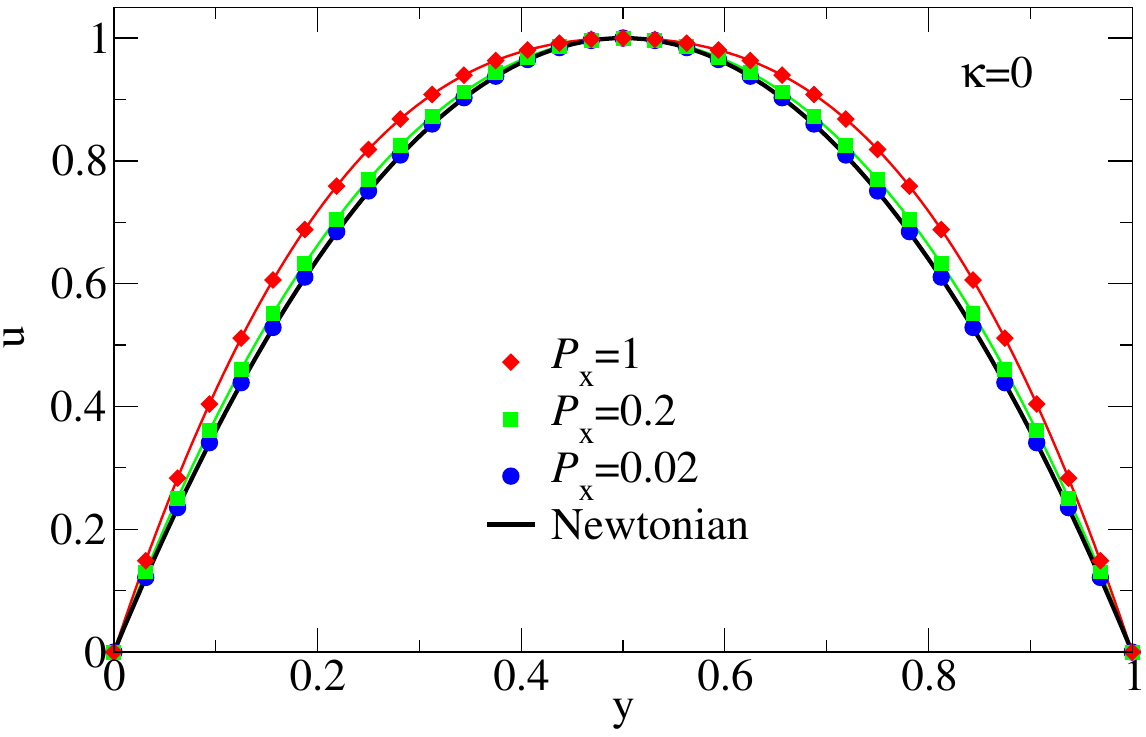} \ \
\includegraphics[width=2.5in, clip]{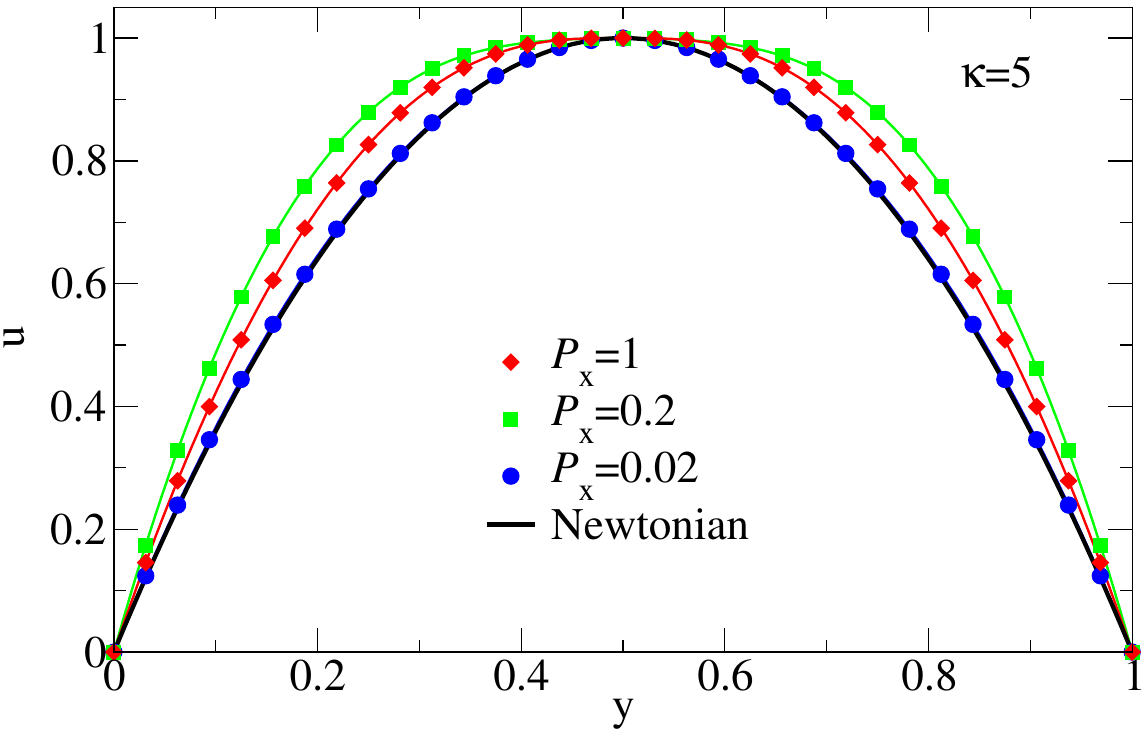}
\caption{
Velocity profiles of flexible and semiflexible oligomer chains in comparison with Newtonian flow profiles.
The velocity profiles shown in Fig.~\ref{fig:uprof} were normalized by the corresponding $\max (u(\cdot,y))$.
}
\label{fig:normu}
\end{figure}

To compare different non-Newtonian  and the Newtonian flows we plot in Fig.~\ref{fig:normu}
the velocity distributions normalized by the maximal velocity together with the Newtonian fluid solution.
(i) At low pressure drop the flow of both flexible and semiflexible melts is Newtonian and it deviates from the
Newtonian regime as the external pressure difference increases.
(ii) The non-Newtonian effect increases progressively
in the flexible chain melt, but it is non-monotone for the semiflexible chains: $P_\textrm{x}=0.2$ causes a
larger non-Newtonian effect on the velocity distribution than $P_\textrm{x}=1$.
The reason for this retrograde non-Newtonian behavior is the following: The semiflexible chain melt is coming
into the regime of a second plateau of viscosity at high shear rates where the melt behaves like
at low shear rates
but the flow velocity is larger due to lower viscosity.
The conjecture on the existence of the second plateau at high shear rates is based in the existence of
positive curvature of
$\eta(\dot\gamma)$ for $\dot\gamma>0.01$ in Fig.~\ref{fig:carreau}.
In contrast, the microscopic data for a flexible chain does not indicate positive curvature within
the measured shear rates.


\begin{figure}[hbtp]
\centering
\includegraphics[width=3.5in, clip]{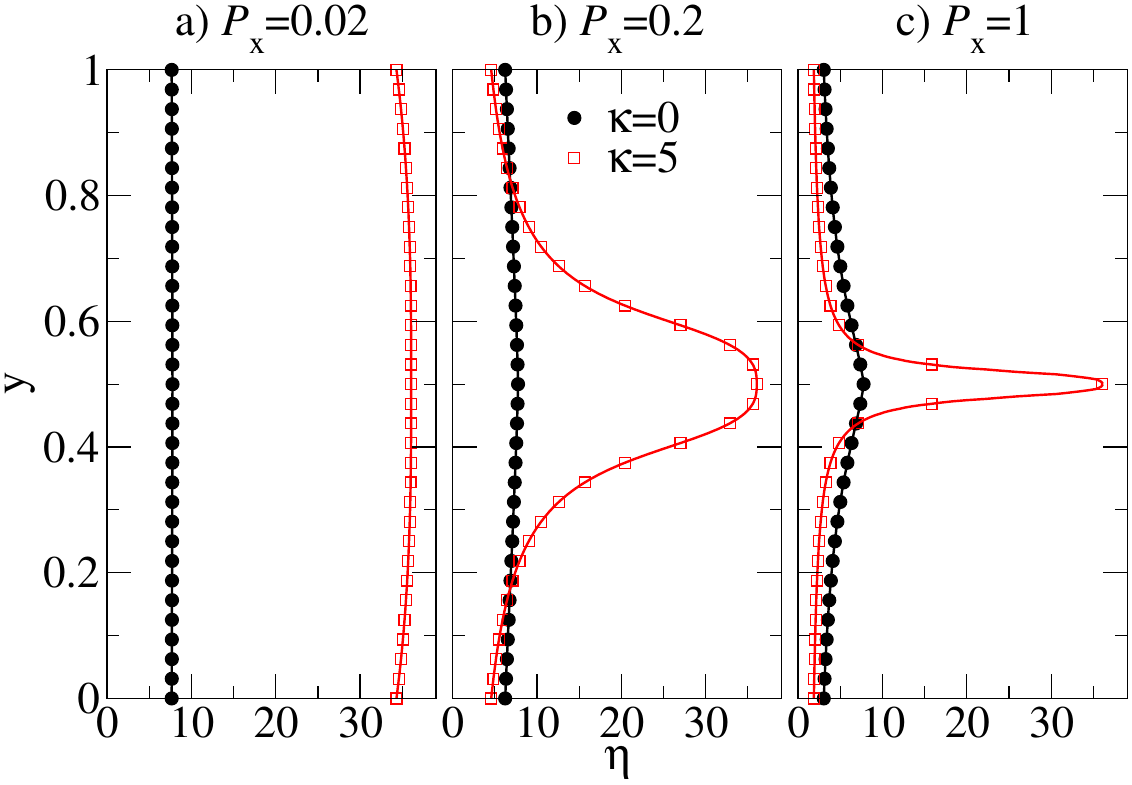}
\caption{
Steady-state viscosity distributions across the channel flow of flexible and semiflexible oligomer melts at various external pressure drops $P_\textrm{x}$.
}
\label{fig:viscprof}
\end{figure}

We proceed with the analysis of viscosity distributions $\eta(y)$, 
 see Fig.~\ref{fig:viscprof}. For slowly flowing melts ($P_\textrm{x}=0.02$ in Fig.~\ref{fig:viscprof}a) the viscosity is almost constant in flexible and semiflexible  chain melts.
At moderate pressure drop $P_\textrm{x}=0.2$, the viscosity is almost constant in the flexible chain melt, but it is $\Lambda$-shaped with a large amplitude variation between the center of the channel and near the walls for the semiflexible  chain melt. At large pressure drop ($P_\textrm{x}=1$), the semiflexible  chains develop a kind of viscosity spike localized at the center of the channel due to $\dot\gamma(\ell/2)=0$. 
Except for this singular region the viscosity distribution of semiflexible  chains is very flat and the viscosity value is smaller  than in the 
flexible chain melt.


 The second test is devoted to the study of  complex geometry effects.  Here we investigate non-Newtonian flows in a channel with a backward-facing step, see Armaly et al.~\cite{armaly}. Newtonian flows in such a channel have been studied in, e.g.,  \cite{florianSIMPLE}.
 Depending on the inflow velocity this flow can be laminar, transitional or turbulent. In the laminar regime at low Reynolds number  one or more recirculation zones of the secondary flow after the expansion arise. One zone is located directly behind the backward-facing step and it can be observed already for a very low Reynolds number. Further zones randomly appear and disappear in the case of high Reynolds numbers. In this test there is  a broad distribution of the shear rates since the intensity of the flow and the shape of streamlines is very different in the main and the secondary flows.

\begin{figure}[htbp]
	\centering
	\includegraphics[width=0.305\textwidth]{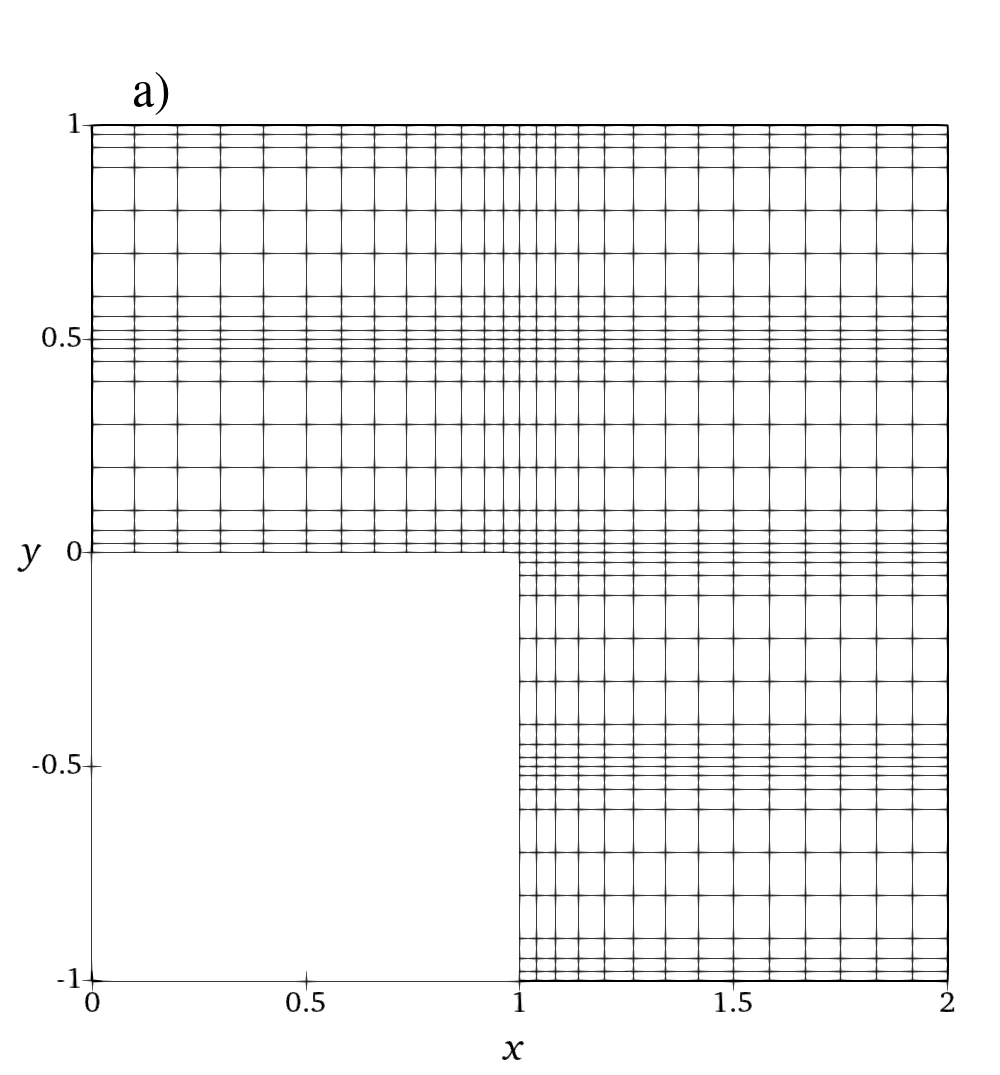}
    \includegraphics[width=0.3\textwidth]{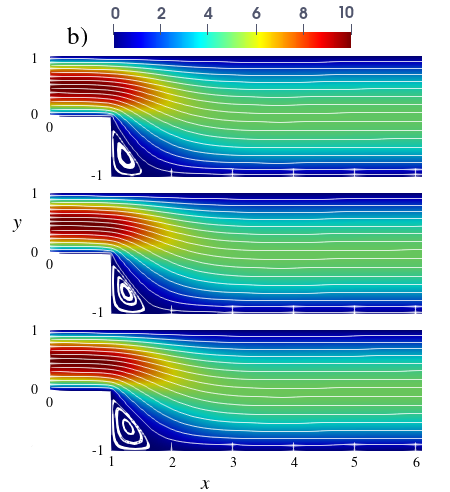}
	\caption{
 	a) Structured quadrilateral mesh used in the Armaly experiment: a cut-out  [0..2]$\times$[-1..1] of the full domain [0..10]$\times$[-1..1] is shown. 
		Mesh resolution increases smoothly at $x\approx 1$ and at $y \in \{-1,-0.5,0,0.5,1\}$. Mesh step varies  between $h_x=1/10$ and $1/26$ for the $x$-direction and between $h_y=1/10$ and $1/48$ for the $y$-direction. 
		b) Velocity and streamlines of the non-Newtonian flows of flexible, semiflexible chain molecules, and of the Newtonian flow, $U_{\rm inlet}=10$, $\kappa=0$ (top),  $\kappa=5$ (middle), $\eta=1$ (bottom).}
    \label{fig:armaly_velU10}
\end{figure}

 We study the flow of chain molecules in the channel of the length $\ell=10$ and the height at the inlet $L_{\rm inlet}=1$ and at the outlet $L_{\rm outlet}=2$. The inlet is located at $x=0$. 
The computation domain is a structured quadrilateral mesh shown in Fig.~\ref{fig:armaly_velU10}a.
 In the $x$-direction the mesh is homogeneous with the mesh step $h_x=1/10$ except for the region near the backward-facing step ($x\approx 1$) where the grid density has been smoothly increased in order to better resolve the transit region. The finest mesh step is $h_x=1/26$. 
 In the $y$-direction the mesh has several high-resolution regions: near the walls at $y \in \{-1,0,1\}$, in the middle of the inlet part ($y=0.5$) and of the main part ($y=0$), and one zone at $y=-0.5$ to keep the mesh symmetry around the central line $y=0$ in the main part of the channel. The finest mesh resolution in these zones is $h_y=1/48$, the coarse resolution is $h_y=1/10$ elsewhere with a smooth transition in-between.
 
 At the inlet  at $x=0$ the Dirichlet  boundary condition is applied for the inflow velocity $(u(0,y)=4U_{\rm inlet}y(1-y), v(0,y)=0)$, where  $U_{\rm inlet}$ is the amplitude of the inflow velocity used as a control parameter for simulations. The outlet boundary conditions at $x=\ell$ 
 impose zero stress (Neumann boundary conditions), i.e. $\grvec{\sigma} \cdot \vec{n} = 0$.

\begin{figure}[htbp]
	\centering
	\includegraphics[width=0.6\textwidth]{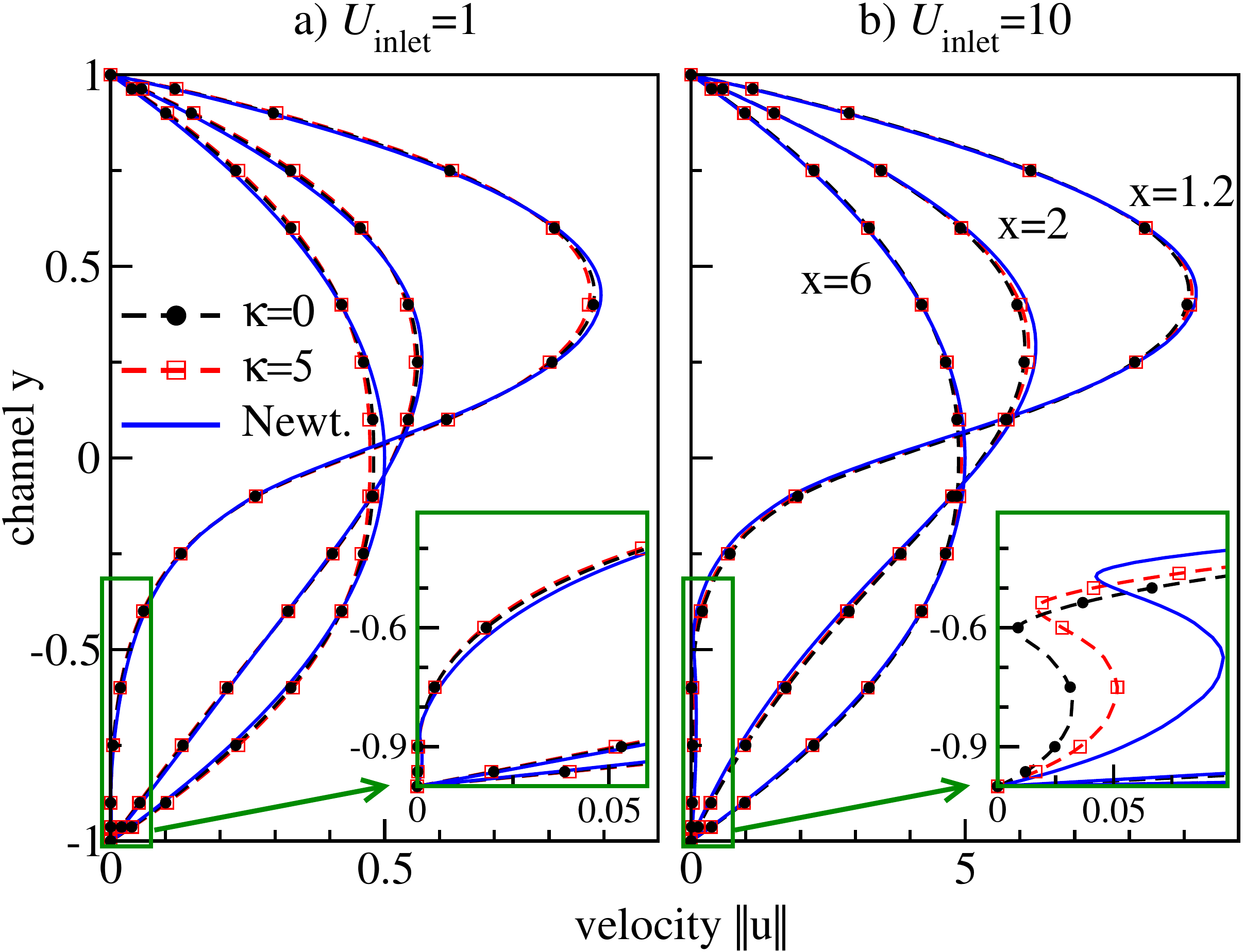}
	\caption{Velocity profiles at  $x \in \{1.2,\, 2,\, 6\}$ of  Newtonian (blue curves) and non-Newtonian flows of flexible $\kappa=0$ (black circles) and semiflexible $\kappa=5$ (red squares) chain molecules, $U_{\rm inlet} \in \{1,10\}$. Insets: zooming into the region of the secondary flow.}
	\label{fig:armaly_prof_U}
\end{figure}

 Figs.~\ref{fig:armaly_velU10}b compares the velocity and the flow streamlines for flexible, semiflexible chain molecules and the Newtonian fluid for inflow velocity $U_{\rm inlet}=10$.
 Although the overall pictures look similar one can clearly observe variations in the recirculation zone in the corner behind the step. To analyze the solution in details we plot in Fig.~\ref{fig:armaly_prof_U} velocity cross sections at $x=1.2,2,6$ for different inflow velocities.
 These positions  represent three characteristic regions of the flow behind the backward-facing step: the secondary flow vortex at the bottom corner directly behind the backward-facing step ($x=1.2$),  flow directly behind the secondary flow ($x=2$), and flow far behind the secondary flow ($x=6$).
 The velocity profiles change between the $x=1.2$ and $x=6$ from strongly asymmetric in the transit region at the backward-facing step to a symmetric one in the region far behind the step, where the characteristics non-Newtonian shape is observed. For $U_{\rm inlet}=1$, the curve for flexible chains is located between the curves for semiflexible chains and Newtonian fluid. For $U_{\rm inlet}=10$, the curves for flexible and semiflexible chains swap indicating the retrograde non-Newtonian effect discussed also for the Poisseuile flow. This effect becomes more visible in the case of high inflow velocity, cf.~recirculation region shown in the insets of Fig.~\ref{fig:armaly_prof_U}.

\begin{figure}[htpb]
	\centering
	\includegraphics[width=0.6\textwidth]{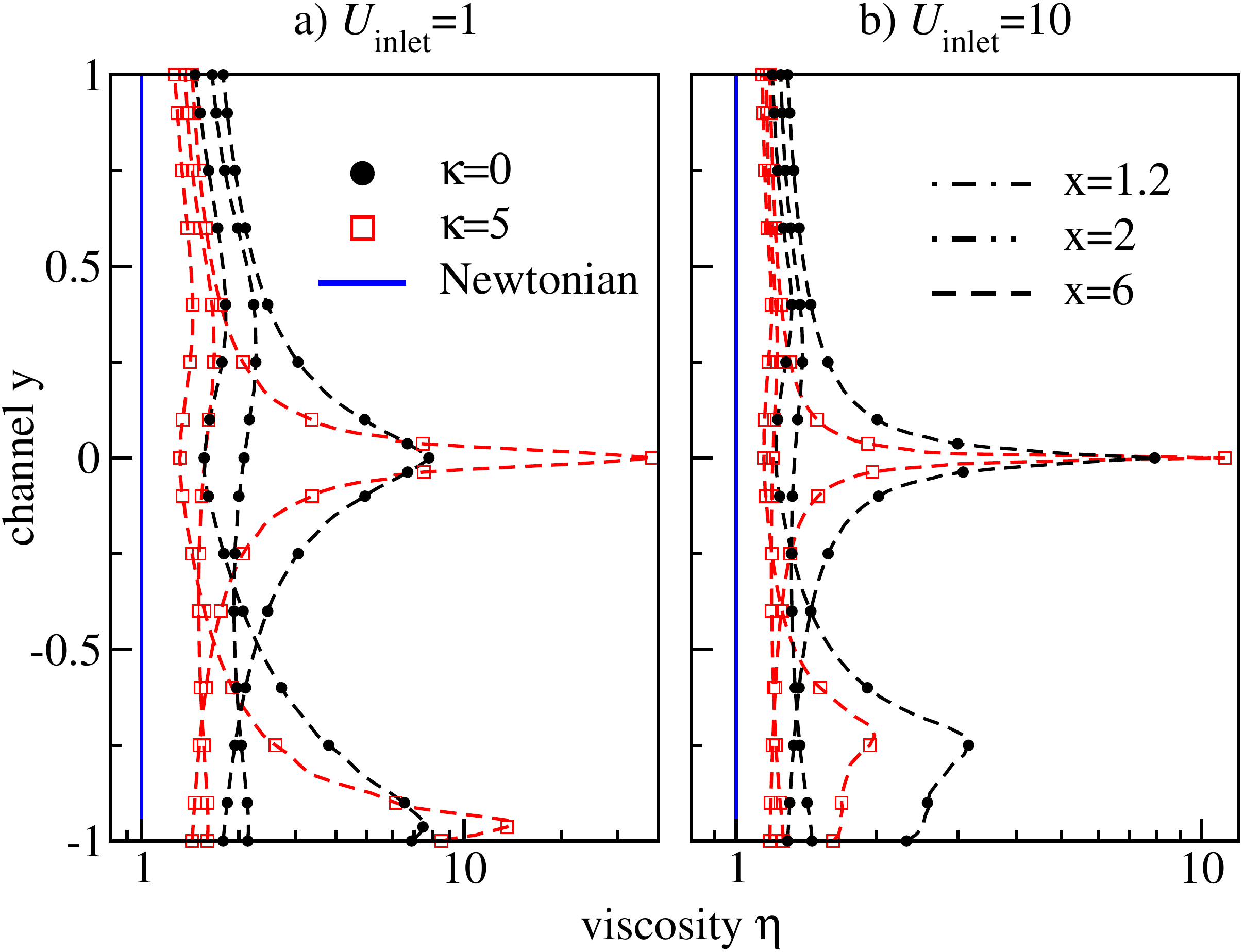}
	\caption{Viscosity profiles at  $x \in \{1.2,\, 2,\, 6\}$ of non-Newtonian flows of flexible (black circles) and semiflexible (red squares) chain molecules, $U_{\rm inlet} \in \{ 1,10 \}$.  Newtonian flow viscosity $\eta=1$ is shown with blue lines.}
	\label{fig:armaly_prof_visc}
\end{figure}

 Fig.~\ref{fig:armaly_prof_visc} shows a detailed analysis of the viscosity profiles at the  cross-sections $x=1.2, \, 2, \, 6$. 
One sharp viscosity peak at $y=0$ is located at the center of the main part of the channel far behind the backward-facing step.  Analogous peak has already been observed for the non-Newtonian Poiseuille flow.
Another peak arises at $y < -0.5$ due to the secondary flow at the corner indicating that the non-Newtonian effect plays an important role in this region, too.

Our extensive simulations explain that (i) the retrograde non-Newtonian behavior is a result of a nearly homogeneous distribution of the viscosity $\eta(y)$ at high shear rates in the flow regime when $\eta(\dot\gamma)$ does not vary significantly (post shear-thinning plateau).
(ii) By tuning the flow parameters it is possible to control the strength of the shear-thinning  effect in a melt by varying the area under the $\Lambda$-shaped viscosity curve in Fig.~\ref{fig:viscprof}.
(iii) A higher flow intensity leads to a wider shear rate distribution and to a narrower $\Lambda$-shaped viscosity spike. 
Note that the height of this spike is limited by the shear viscosity in the non-sheared system as the flat region of the velocity profile corresponds to shear rates close to zero.

\section{Summary and Outlook}
\label{sec_6}

In this manuscript we have investigated and reviewed concisely the molecular origins of shear-thinning at moderate to high shear rates in a high-density, low molecular weight melt of coarse-grained bead-spring polymers as a function of chain length and stiffness. While zero shear rate viscosity for flexible chains \cite{Kroger_1993} and the accompanying tendency towards shear-thinning increases linearly with chain length, the course of viscosity at fixed chain length becomes non-monotone if stiffness is taken into account. Although at high shear rate viscosity decreases with increasing stiffness and drops even below the corresponding values for monomers, at low shear rates and in equilibrium viscosity first increases as a function of stiffness before decreasing for stiff chains \cite{Xu_2017}.While the initial rise of viscosity can be attributed to the emergence of entanglements in semiflexible chains, the following decline can be explained by a collective alignment of chains (in the context of an isotropic-nematic transition in equilibrium) which is further amplified in the presence of shear. The alignment is, of course, not present in simulations of single chains and as such expected to vanish gradually with decreasing density.

We also identified progressive alignment of chains with shear rate as a major cause for shear-thinning \cite{Kroger_1997,Xu_2017} and studied movement modes of individual chains. While in the flexible case, 
individual chains tumble at moderate and high shear rates, for semiflexible chains we observe a transition from stretched conformations at moderate rates to a state in which stretched and tumbling chains coexist in agreement with \cite{Kong2019}. The latter leads to a small decrease of average chain sizes at high rates and surprisingly does not impede the continuous trend towards lower viscosities. Both alignment and the emergence of tumbling can also be observed qualitatively in simulations of single chains under shear and are thus expected to occur also in dilute and semi-dilute solutions.

In Section~\ref{sec_5} we have demonstrated the influence of microscopic shear-thinning behaviour on macroscopic channel flow. Using the hybrid multiscale method that couples Molecular Dynamics and discontinuous Galerkin schemes we have studied flows of oligomer melts in two types of  channels for flexible and semiflexible  molecules, respectively. We have observed a transition from a Newtonian to a non-Newtonian flow (as well as a retrograde non-Newtonian effect for semiflexible chains) caused by the shear-thinning viscosity effects. Furthermore,  we have also analysed the effects of complex geometry. In the near future we would like to investigate shear-thinning in two-component melts consisting of polymers with varying lengths and stiffnesses as well as consequences of the latter for the macroscopic flow in simple and complex geometries.

\section*{Acknowledgements}
We are grateful to the Deutsche Forschungsgemeinschaft (DFG, German Research Foundation) for funding this research: Project number 233630050-TRR 146. (projects C1, C5 and C7). The authors gratefully acknowledge
the computing time granted on the HPC cluster
Mogon at Johannes Gutenberg University Mainz.

\section{References}
\bibliographystyle{unsrt}
\bibliography{cit1}

\appendix

\section{Alternative Time-scale Mapping}
In addition to the simple time scale mapping between single chain simulations and melts described at the end of section 2, here, we consider matching Rouse times for each value of $\kappa$. Note that while this procedure works well for $\kappa\leq6$ and $\kappa=10$, we did not get a convincing exponential fit for the autocorrelation function of the sheared melt in the intermediate range of stiffnesses, likely due to the presence of entanglements. Excluding values for $\kappa=7,8$ and $9$, the relative scaling factor $\zeta(\kappa)/\zeta(\kappa=0)$ was 1.15 for $\kappa=1$, 1.29 for $\kappa=2$, 1.46 for $\kappa=3$, 1.81 for $\kappa=4$, 2.07 for $\kappa=5$, 2.58 for $\kappa=6$, and finally, 2.52 for $\kappa=10$.  The results reported in Sec.~\ref{sec_3}, for the most part do not change qualitatively in comparison to the simplified matching.

\begin{figure}[h!]
\centering
\includegraphics[width=3.5in,clip]{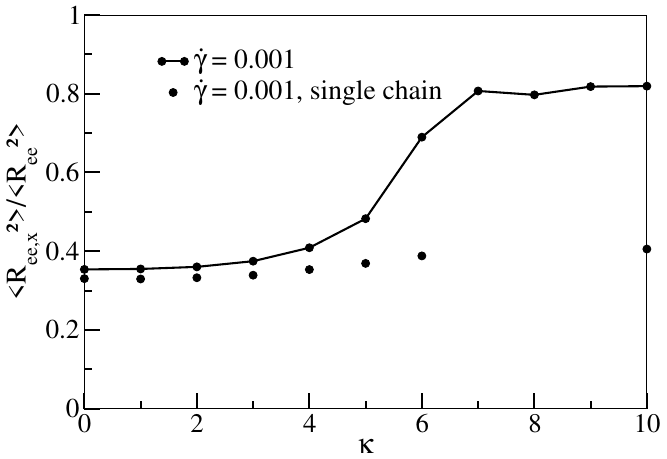}]
\caption{$\langle R_{ee,x}^2\rangle / \langle R_{ee}^2\rangle$ as a function of $\kappa$ at $\dot{\gamma} = 0.001$ for melt (line with dots) and single chain simulations (only dots) with the more involved mapping described in the appendix.}

\label{A1}
\end{figure}

If we apply this time-scale matching procedure to the results shown in Figs.~\ref{fig2}a and \ref{fig2}b, the single chain curves corresponding to $\kappa= 5$ and $10$ will be shifted to the left by factors of $2.07$ and $2.52$, respectively, which will not change our qualitative conclusions. Fig.~\ref{A1} (which corresponds to Fig.~\ref{fig4}b) shows $\langle R_{ee,x}^2\rangle / \langle R_{ee}^2\rangle$ as a function of $\kappa$ at $\dot{\gamma} = 0.001$. For single chain simulations, the ratio is no longer flat, but  rises marginally for larger values of $\kappa$. Again, this outcome does not change the qualitative conclusions described in the main text.
\end{document}